\documentclass[amsmath,amssymb,amsfonts,aps,pre,preprint,superscriptaddress,bibnotes,showpacs,showkeys,longbibliography]{revtex4-1}
\usepackage{mathtools}
\usepackage{epsfig}
\usepackage[english]{babel}
\usepackage{physics}
\usepackage{color}
\usepackage{subcaption}
\usepackage{hyperref}

\newcommand*{\logten}{\mathop{\log_{10}}}
\newcommand*{\lop}{\mathcal{L}_z}
\newcommand*{\nlop}{\mathcal{N}_z}
\newcommand*{\rop}{\mathcal{R}_z}

\AtBeginDocument{%
    \newwrite\bibnotes
    \def\bibnotesext{Notes.bib}
    \immediate\openout\bibnotes=\jobname\bibnotesext
    \immediate\write\bibnotes{@CONTROL{REVTEX41Control}}
    \immediate\write\bibnotes{@CONTROL{%
    apsrev41Control,author="08",editor="1",pages="1",title="0",year="1"}}
     \if@filesw
     \immediate\write\@auxout{\string\citation{apsrev41Control}}%
    \fi
}%

\begin{document}
\title{BLUES iteration applied to nonlinear ordinary differential equations for wave propagation and heat transfer}

\author{Jonas Berx}
\affiliation{Institute for  Theoretical Physics, KU Leuven, B-3001 Leuven, Belgium}

\author{Joseph O. Indekeu}
\affiliation{Institute for  Theoretical Physics, KU Leuven, B-3001 Leuven, Belgium}

\date{\today}

\begin{abstract}
The iteration sequence based on the BLUES (Beyond Linear Use of Equation Superposition) function method for calculating analytic approximants to solutions of nonlinear ordinary differential equations with sources is elaborated upon. Diverse problems in physics are studied and approximate analytic solutions are found. We first treat a damped driven nonlinear oscillator and show that the method can correctly reproduce oscillatory behaviour. Next, a fractional differential equation describing heat transfer in a semi-infinite rod with Stefan-Boltzmann cooling is handled. In this case, a detailed comparison is made with the Adomian decomposition method, the outcome of which is favourable for the BLUES method. As a final problem, the Fisher equation from population biology is dealt with. For all cases, it is shown that the solutions converge exponentially fast to the numerically exact solution, either globally or, for the Fisher problem, locally. 
\end{abstract}

\maketitle

\section{Introduction}
Differential equations (DEs) are the cornerstones of modern physics and mathematics. While linear DEs are quite well understood, nonlinear DEs remain elusive objects. Notwithstanding this, most physical systems exhibit some form of nonlinearity. It is a challenge to obtain exact analytical solutions whenever possible. Some analytical or semi-analytical techniques such as the Adomian decomposition method (ADM), the homotopy analysis method (HAM) or perturbative techniques such as the soliton perturbation theory have been proposed and proved to be extremely useful \cite{adomian,ham,karpman1977,Keener}. When an external force is also present in the form of an inhomogeneous source or sink \cite{MAEDA,Ermakov2019}, one often turns to numerical solutions. These solutions do not offer the richness of information that analytical solutions can provide. While linear DEs with sources can be treated by the theory of Green functions and the principle of superposition, nonlinear DEs violate the superposition principle and a solution cannot be constructed in this way. In this paper, following up on our Letter \cite{Berx}, we investigate further how the practice of Green functions can be usefully extended to nonlinear DEs with an inhomogeneous source or sink, effectively using the superposition principle beyond the linear domain. 

Relative to previous works on this method our contribution is situated as follows. In \cite{smets} the usefulness was demonstrated of simple exponential tail solutions of nonlinear reaction-diffusion-convection DEs describing traveling wave fronts. These simple solutions are exact provided a co-moving Dirac delta function source is added to the DE. In \cite{BLUES} it was observed that for some problems one and the same simple exponential tail solution simultaneously solves the nonlinear DE with a Dirac delta source as well as a related linear DE with the same Dirac delta source. This observation led to the idea to formulate an analytic method that uses the concept of Green function, and its convolution with an arbitrary source, beyond the linear domain. This approach was named BLUES (Beyond Linear Use of Equation Superposition) and it was suggested that the method can be useful as a  perturbation expansion, the small parameter being the ratio of the width of the source to the decay length of the tail. Subsequently, in \cite{Berx} quantitative calculations were performed, demonstrating that the method entails an analytic iteration procedure which converges exponentially rapidly for a variety of problems, and which is non-perturbative. There is no need for a small parameter. Furthermore, the method was also used the other way around, starting from a linear DE and freely adding a nonlinearity to it, instead of starting from a nonlinear DE and looking for a related linear DE. Applications were presented to solitary wave solutions of the  Camassa-Holm equation and traveling wavefront solutions of the Burgers equation. Our contribution now is to extend substantially the applicability of the method and demonstrate its accuracy for problems involving oscillatory waves, phenomena described by fractional differential equations (FDEs), and systems with nonlinear growth. We present a detailed comparison with an alternative method, the ADM. In the new applications we show that the convergence of the method is sometimes local instead of the global convergence illustrated in \cite{Berx}. The applications until now were for problems that can be reduced to an ordinary DE with a co-moving source. This limitation is henceforth removed.

This paper is organized as follows. In Section \ref{sec:BLUES} we recall briefly the analytic iteration procedure in order to make the paper sufficiently self-contained. In Section \ref{sec:oscillator}, we study the damped nonlinear oscillator and show that the method allows one to approximate correctly the decaying oscillations. In Section \ref{sec:fractional}, a model for heat transfer in a semi-infinite rod with a fractional-order derivative is treated, and employed to show that the BLUES function method can be extended to the arena of fractional differential equations. Also in this Section we compare the performance of the method to that of the ADM. Finally, in Section \ref{sec:fisher}, a wave solution of a nonlinear extension of the heat equation involving reaction, i.e., the Fisher equation, is treated. We close the paper with our conclusions and outlook.

\section{The BLUES iteration method}
\label{sec:BLUES}
Here we recapitulate concisely the BLUES function concept \cite{BLUES} from the viewpoint of the efficient iteration procedure that was developed from it in \cite{Berx}. One starts from a linear ordinary DE which can be written as an operator $\lop$ acting on a function $U(z)$ and assumes that one knows the (piecewise analytic) Green function $U(z) = G(z)$ which solves
\begin{equation}
    \label{eq:linear_operator_delta}
    \lop U(z) = \delta(z),
\end{equation}
with suitable boundary conditions. The Dirac delta source compensates a possible discontinuity of the derivative of order $n-1$ at $z=0$ in the case of an $n$-th order DE.  Next one  considers the linear DE with an arbitrary source $\psi(z)$. The solution is then the convolution product $G\ast\psi$.

One may add a nonlinearity rather freely, but so that the same boundary conditions are respected, and arrive at the nonlinear ordinary DE
\begin{equation}
    \label{eq:nonlinear_DE_psi}
    \nlop U(z) = \psi(z),
\end{equation}
A solution is then proposed in the form $U(z) = B\ast\phi$. The function $B(z)$ is called BLUES function and it is taken to be the Green function of the linear DE, so $B(z) = G(z)$. The challenge is to calculate the associated source $\phi(z)$ for the given source $\psi (z)$, knowing that $B\ast\psi$ solves the linear DE with source $\psi(z)$. For achieving this one defines a residual operator $\rop \equiv \lop - \nlop$ and makes use of the implicit identity
\begin{equation}
    \label{eq:phi}
    \phi(z) = \psi(z) + \rop(B\ast\phi)(z).
\end{equation}

To obtain the solution to the nonlinear DE \eqref{eq:nonlinear_DE_psi}, equation \eqref{eq:phi} can be iterated in order to calculate an approximation in the form of a sequence in powers of the residual $\rop$. To zeroth order, the sources $\phi^{(0)}(z)$ and $\psi(z)$ are identical and the approximation is the convolution product
\begin{equation}
    \label{eq:zeroth_order}
    U_\psi^{(0)}(z) = (B\ast\phi^{(0)})(z)= (B\ast\psi)(z).
\end{equation}
To $n$th order ($n\geq1$), the approximate solution can be found by iterating \eqref{eq:phi} and taking the convolution product with $B(z)$ \cite{Berx}
\begin{equation}
    \label{eq:nth_order}
    U_\psi^{(n)} = (B\ast\phi^{(n)})(z) =  U_\psi^{(0)}(z) + \left(B\ast\rop U_\psi^{(n-1)}\right)(z).
\end{equation}
We now proceed to applications beyond what was presented in \cite{Berx}.

\section{The damped nonlinear oscillator}
\label{sec:oscillator}
We start from a general linear wave equation in one dimension with a co-moving Dirac delta source with reduced amplitude $s$
\begin{equation}
    \label{eq:wave_pde}
    u_{tt} - u_{xx} + \gamma u_x + u = s\, \delta(x-ct),
\end{equation}
in which the displacement $u$, time $t$, space $x$, amplitude $\gamma$ and velocity $c$ are also reduced so as to be dimensionless,
and look for traveling wave solutions by transforming to the coordinate $z = x-ct$ and restricting $U(z) \equiv u(x,t)$,
\begin{equation}
    \label{eq:wave_linear}
    \alpha U_{zz} + \gamma U_z + U = s\, \delta(z),
\end{equation}
where $\alpha = c^2 -1$. Derivatives (with respect to $t$, $x$ or $z$) have been denoted by subscripts.

This DE is solved by the Green function, to be used as BLUES function,
\begin{equation}
    \label{eq:wave_BLUES}
    B(z)\equiv\begin{cases}
        0, & z<0 \\ 
        \sin{\left(\frac{\lambda z}{2\alpha}\right)}\mathrm{e}^{-\frac{\gamma z}{2\alpha}}, & z\geq0,
    \end{cases}
\end{equation}
with $\lambda = \sqrt{4\alpha - \gamma^2}$ and source amplitude $s = \lambda/2$. Now an arbitrary nonlinear term can be added, which we choose to be the cubic-quintic function $\beta U^3 + \xi U^5$, where $\beta$ and $\xi$ are tuneable parameters. Altogether the nonlinear wave equation with an arbitrary source is
\begin{equation}
    \label{eq:wave_nonlinear}
    \alpha U_{zz} + \gamma U_z + U + \beta U^3 + \xi U^5= s \psi(z),
\end{equation}
where again the amplitude is $s = \lambda/2$ and the 1-norm (i.e., the integral over $z$) of the source $\psi(z)$ is unity. This DE is a basic model for a myriad of physical systems. When $\beta$ and $\xi$ are chosen to be $-1/3!$ and $1/5!$ respectively, the terms $U + \beta U^3 + \xi U^5$ can be interpreted as the first three nonzero terms in the sine Taylor series. Including higher-order terms in the series, one can construct the nonlinear DE
\begin{equation}
    \label{eq:sine-gordon}
    \alpha U_{zz} + \gamma U_z + \sin{U}= s \psi(z),
\end{equation}
which is an equation for the damped and driven Sine-Gordon model \cite{Ekomasov2018}, often used to describe the dynamics of Josephson junctions in superconductors \cite{Gul,starodub}. Another important application of equation \eqref{eq:wave_nonlinear} is the cubic-quintic Duffing oscillator, which is used to describe damped harmonic motion in a nontrivial potential and has become a paradigm for the study of chaos. For computational convenience we will only include the terms in the sine Taylor series up to and including third order, so we will choose $\xi = 0$.

Following the procedure outlined in Section \ref{sec:BLUES}, we extract the residual operator and identify its action on $U$,
\begin{equation}
    \label{eq:wave_residual}
    \rop U \equiv -\frac{\beta}{s}U^3
\end{equation}
and use this to calculate higher-order approximants using  $B(z)$  in the iteration sequence \eqref{eq:nth_order}. Note that $\rop B (z) \neq 0$. For the source $\psi$ we can choose, e.g., an exponential corner source which  possesses a tunable dimensionless decay length $K$,
\begin{equation}\label{eq:source}
    \psi(z) = \frac{\mathrm{e}^{-|z|/K}}{2K}
\end{equation}
The choice of the source \eqref{eq:source} is not limited to functions with non-zero 1-norm; we have also performed calculations for sources with 1-norm zero and found that this has no significant impact on the convergence or accuracy of the method. 

The zeroth-order approximant for the solution of \eqref{eq:wave_nonlinear} can be calculated by performing the convolution integral \eqref{eq:zeroth_order} with BLUES function \eqref{eq:wave_BLUES} and normalized exponential corner source \eqref{eq:source}

\begin{equation}
    \label{eq:wave_conv}
   U^{(0)}_{\psi}(z) = \begin{cases}
        \frac{K\lambda}{4C_+}\mathrm{e}^{z/K}, & z<0 \\ 
        \left[A \sin{\left(\frac{\lambda z}{2 \alpha}\right)} - B\lambda\cos{\left(\frac{\lambda z}{2\alpha}\right)}\right]\frac{\mathrm{e}^{-\frac{\gamma z}{2\alpha}}}{2} + \frac{K\lambda}{4C_-}\mathrm{e}^{-z/K}, & z\geq0,
    \end{cases}
\end{equation}
where the constants $A,B,C_\pm$ are introduced to simplify notation. They are given by combinations of $\alpha$, $\gamma$ and $K$:
\begin{equation}
    \begin{split}
        A &= \frac{2\alpha^2 - K^2\gamma^2 + 2\alpha K^2}{\alpha^2 - K^2\gamma^2 + 2\alpha K^2 + K^4} \\
        B &= \frac{K^2\gamma}{\alpha^2 - K^2\gamma^2 + 2\alpha K^2 + K^4} \\ 
        C_\pm &= \alpha \pm K\gamma + K^2.
    \end{split}
\end{equation}
Higher orders can in principle be calculated using equation \eqref{eq:nth_order} but in practice this is not feasible without the aid of mathematical software.

\begin{figure}[!ht]
    \centering
    \includegraphics[width = 0.85 \linewidth]{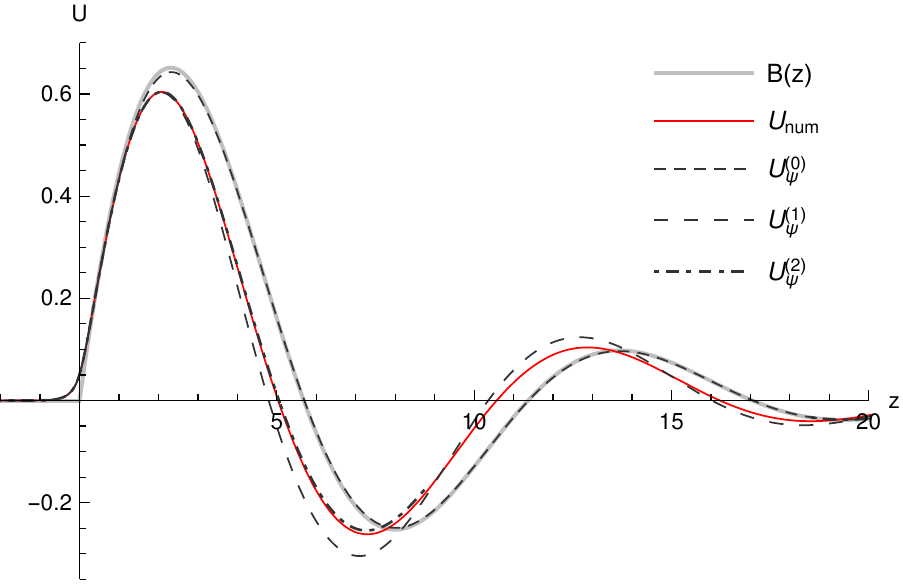}
 \caption{ Traveling wave solution to the nonlinear wave equation \eqref{eq:wave_nonlinear} with an exponential corner source \eqref{eq:source}. The numerical solution $U_{\rm num}$ (red, full line), the zeroth-order $U_\psi^{(0)}$ (black, dashed line), the first-order approximant $U_\psi^{(1)}$ (black, wider spaced dashed line) and the second-order approximant $U_\psi^{(2)}$ (black, dot-dashed line, calculated for $z<9$) are compared. The BLUES function (gray, solid line) is also shown. Parameter values are $\alpha = 3$, $\gamma = 1$, $\beta = 1$, $\xi = 0$ and $K = 1/5$.}
 \label{fig:W_k02}
\end{figure}
\begin{figure}[!ht]
\centering
    \includegraphics[width = 0.85\linewidth]{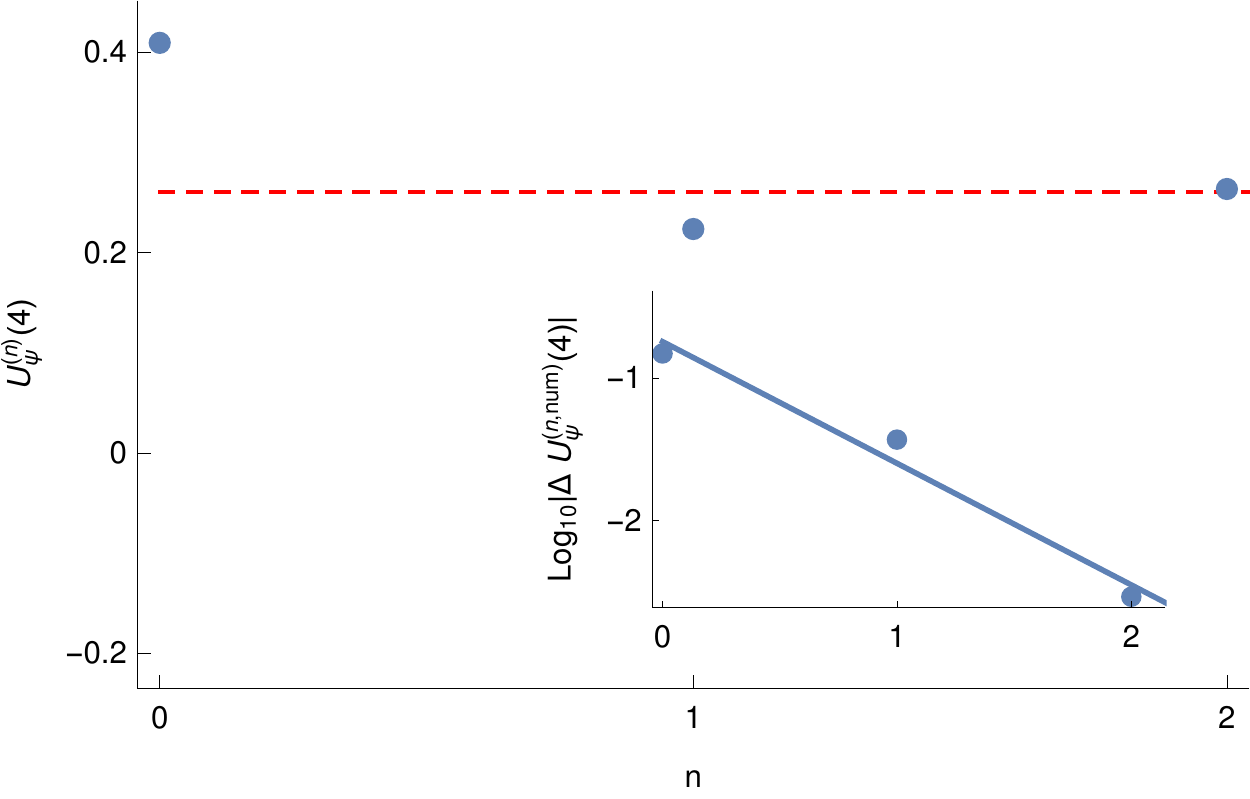}
\caption{Wavefront values $U_\psi^{(n)}(z=4)$  versus order $n$ for the nonlinear wave equation \eqref{eq:wave_nonlinear}. The numerically exact value (red, dashed line) is also shown. \textbf{Inset:} A $\logten$ semi-log plot of the increments  $|\Delta U_\psi^{(n,num)}(4)|\equiv |U_\psi^{(n)}(4) - U^{(num)}(4)|$ of the approximants versus $n$, and a linear fit. Parameter values are $\alpha = 3$, $\gamma = 1$, $\beta = 1$, $\xi = 0$ and $K = 1/5$.}
 \label{fig:W_max_k02}
\end{figure}
\begin{figure}[!ht]
    \centering
    \includegraphics[width = 0.85\linewidth]{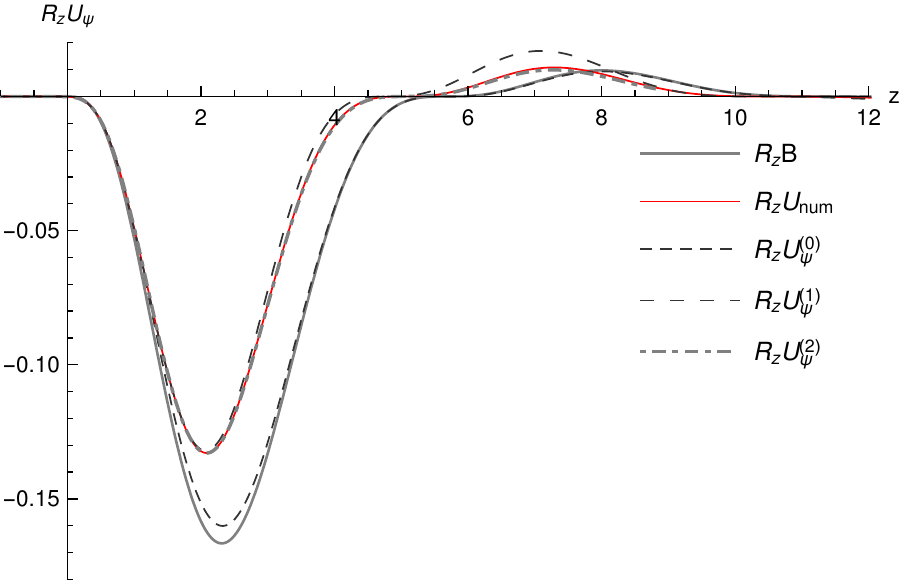}
\caption{Residual function $\rop U_\psi^{(n)}$ of the approximants ($n=0,1,2$) for the nonlinear wave equation \eqref{eq:wave_nonlinear} with exponential corner source \eqref{eq:source}. The residual operator applied to the BLUES function and to the numerical solution (red, full line) are also shown. Parameter values are $\alpha = 3$, $\gamma = 1$, $\beta = 1$, $\xi = 0$ and $K = 1/5$.}
\label{fig:W_res_k02}
\end{figure}

In Fig. \ref{fig:W_k02}, a comparison between the numerically exact solution and  the zeroth-, first- and second-order approximants is made. Note that the zeroth-order approximant almost coincides with the BLUES function (also shown), for $z>0$, while the second-order approximant is on top of the numerically exact solution. Next, we analyse the convergence of the iteration sequence to the exact solution. In Fig. \ref{fig:W_max_k02} the values of the approximants at $z = 4$ for orders $n = 0,1$ and 2 are shown. The inset shows that the convergence to the exact solution is exponentially fast. 

Special attention is given to the calculation of the residual, at iteration $n$, which provides another means of monitoring the convergence of the method. The residual in the BLUES function and the residual functions for $n=0, 1$ and $2$ are shown in Fig. \ref{fig:W_res_k02}, together with the residual operator \eqref{eq:wave_residual} applied to the numerically exact solution (red, full line). Note that the residual functions are negative in the domain around the global maximum of $U(z)$, where $U>0$, that they are zero (with a cubic dependence on $z$) where the curves of Fig. \ref{fig:W_k02} change sign, and that they are positive in the domain around their first minimum, where $U<0$. This is obvious in view of the simple form \eqref{eq:wave_residual}.

In the limit $K \rightarrow 0$, the chosen source converges to the Dirac delta source used to calculate the BLUES function \eqref{eq:wave_BLUES}. Because the BLUES function does not solve \eqref{eq:wave_nonlinear}, also not in this limit, the iteration sequence converges (exponentially fast) to a nontrivial function. One can calculate the first two terms  by setting $\psi(z) = \delta(z)$ in equations \eqref{eq:zeroth_order} and \eqref{eq:nth_order}. To zeroth order, the convolution is identical to the BLUES function. To first order in the iteration, the solution with a Dirac delta function source is given by
\begin{equation}
    \label{eq:wave_delta}
    U_\delta^{(1)}(z) = B(z) - \frac{\beta}{s}\int\limits_{-\infty}^\infty B(z-z_0)B(z_0)^3 \mathrm{d}z_0.
\end{equation}
Note that this does not provide the first-order correction in a series expansion in the parameter $\beta$ because higher-order iterations generate additional contributions linear in $\beta$.

\section{Fractional heat transfer equation}
\label{sec:fractional}

In this section we consider the following nonlinear ordinary fractional differential equation (FDE) for $U(t)$ defined on the semi-infinite real line $t\in [0,\infty)$ with differential order $0<\alpha\leq1$ and exponent $n\geq1$ and with initial condition $U(t=0)= C_0$, with $C_0\geq0$.
\begin{equation}
    \label{eq:nonlinear_FDE}
    \mathcal{N}_t U = D^\alpha_t U + U^n = \psi(t)\, ,
\end{equation}
where $D^\alpha_t$ is the Riemann- Liouville fractional derivative defined as follows, for $\alpha >0$ and $t>0$,
\begin{equation}
    \label{eq:riemann-liouville_fde}
    D^\alpha_t f(t) \equiv \begin{dcases}
        \frac{1}{\Gamma(m-\alpha)}\frac{d^m}{dt^m}\int_0^t \frac{f(\tau)}{(t-\tau)^{\alpha+1-m}}\mathrm{d}\tau, & m-1<\alpha<m\in\mathbb{N}\\
        \frac{d^m}{dt^m}f(t), & \alpha = m\in\mathbb{N}
    \end{dcases}
\end{equation}
where $\Gamma(.)$ is the gamma function. This equation has previously been studied in the context of nonlinear heat transfer for the case that $\alpha = 1/2$, $C_0 = 0$ and $n = 4$ (Stefan-Boltzmann cooling) \cite{wazwaz2012,wazwaz1996}. The calculations that follow are valid for all values of $0<\alpha\leq1$ and $n\geq1$. It has been shown that if $\psi(t)$ is a piecewise continuous bounded function, equation \eqref{eq:nonlinear_FDE} is guaranteed to have a unique solution. If $\psi(t)$ is nondecreasing in an interval $0<t<s$, $s\in(0,\infty)$ then the solution is also nondecreasing in that interval \cite{padamavally,keller1972}. Note that the differential order can in principle be higher than $\alpha = 1$. One can then separate the order $\alpha = m +\beta$ in an integer part $m\in\mathbb{N}$ corresponding to a regular integer-order differential operator, and a fractional part $0<\beta\leq1$ which again corresponds to a fractional differential operator. The FDE \eqref{eq:nonlinear_FDE} should consequently be supplemented with additional boundary (or initial) conditions up to a number $m + 1$. In the remainder of this work, we will assume $0<\alpha\leq1$.

Following the steps in the BLUES procedure outlined in Section \ref{sec:BLUES}, we can simply start from the linear DE obtained by dropping the nonlinear $U^n(t)$ term. We write the linear FDE in operator form
\begin{equation}
    \label{eq:linear_FDE}
    \mathcal{L}_t U = D^\alpha_t U = \psi(t)\, ,
\end{equation}
with arbitrary source $\psi(t)$ and the initial condition chosen to be $U(0) = 0$. The Green function for  \eqref{eq:linear_FDE} can now be calculated by considering a Dirac delta function source instead of $\psi(t)$ 
\begin{equation}
    \label{eq:linear_FDE_Green}
    D^\alpha_t G(t,t') = \delta(t-t')\, ,
\end{equation}
where $t-t' \geq 0$ because the problem is formulated on the semi-infinite real line. This Green function is readily calculated to be
\begin{equation}
    \label{eq:Green_FDE}
    G(t,t') = \frac{(t-t')^{\alpha-1}}{\Gamma(\alpha)}
\end{equation}
Consequently the solution of the linear FDE \eqref{eq:linear_FDE} is the convolution integral of the Green function \eqref{eq:Green_FDE} and the source $\psi(t)$, which we will choose from now on to be the constant function $\psi(t) = 1$ for $t\geq0$, as was done in references \cite{wazwaz1996,wazwaz2012}. 

It is worth emphasizing that this application to heat transfer is fundamentally different from the other ones considered in this paper as well as in \cite{Berx} in that the source is not assumed to be originating from a disturbance that is ``co-moving" with the solution. The variable here is time and not a co-moving coordinate, and the source arises as a natural physical ingredient of the problem. Therefore this example constitutes a non-trivial extension of the domain of applicability of the method not only in the type of DE (from DE to FDE) but also in the character and interpretation of the source term in the DE.

\begin{figure}[!htp]
\begin{subfigure}{\linewidth}
    \centering
    \includegraphics[width = 0.85 \linewidth]{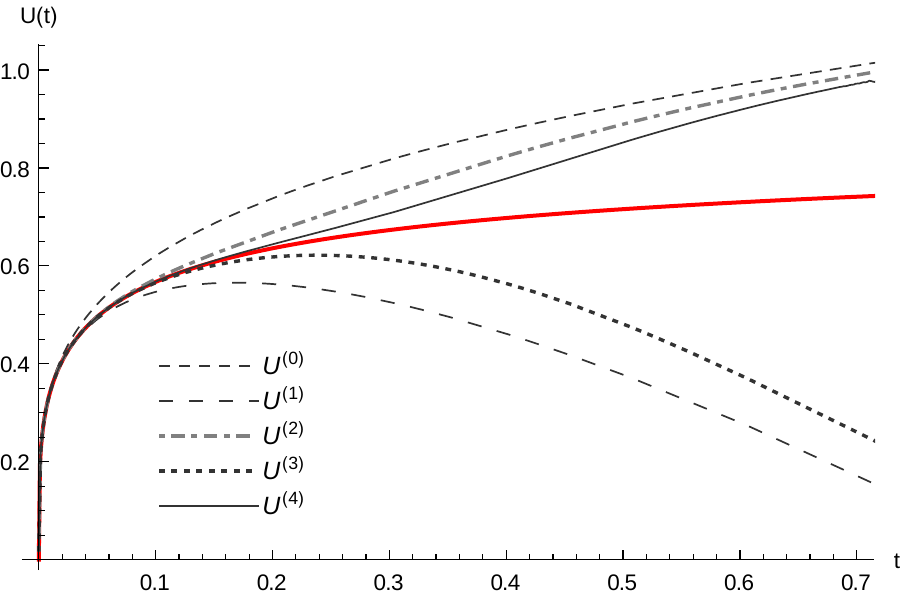}
    \caption{}
    \label{fig:F_quarter}
\end{subfigure}
\begin{subfigure}{\linewidth}
    \centering
    \includegraphics[width = 0.85 \linewidth]{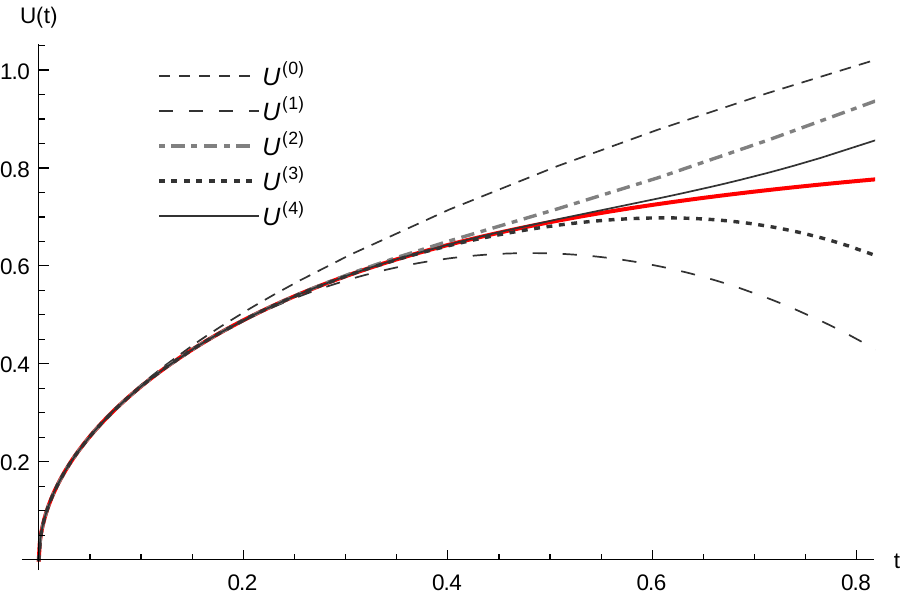}
    \caption{}
    \label{fig:F_half}
\end{subfigure}
\caption{Solution to the nonlinear FDE \eqref{eq:nonlinear_FDE} with constant source $\psi(t) = 1$ and fractional order \textbf{(a)} $\alpha = 1/4$ and \textbf{(b)} $\alpha = 1/2$. The numerical solution (red, full line) is compared with the approximants up to fourth order (black/gray lines).}
\label{fig:F_quarter_half}
\end{figure}

\begin{figure}[!htp]
\begin{subfigure}{\linewidth}
    \centering
    \includegraphics[width = 0.85 \linewidth]{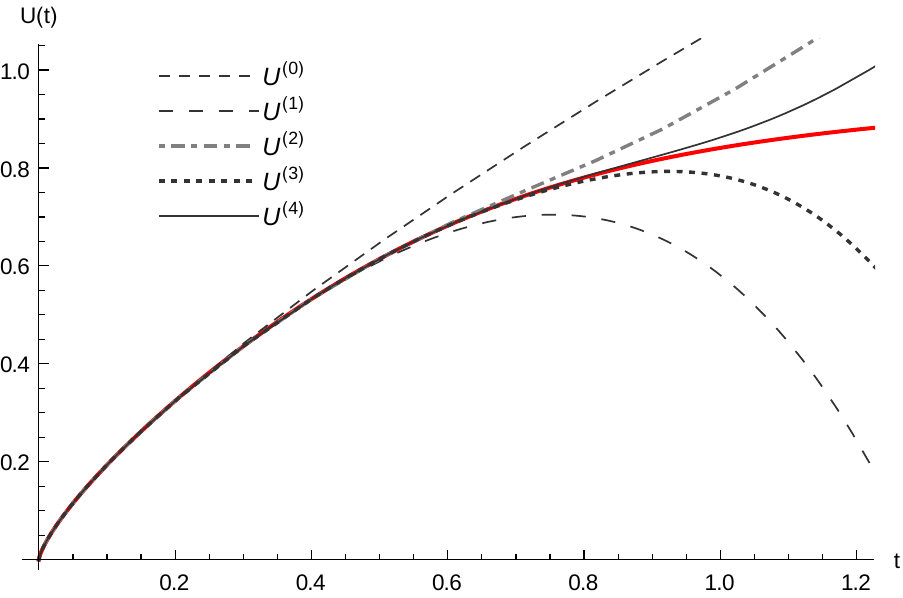}
    \caption{}
    \label{fig:F_threequarter}
\end{subfigure}
\begin{subfigure}{\linewidth}
    \centering
    \includegraphics[width = 0.85 \linewidth]{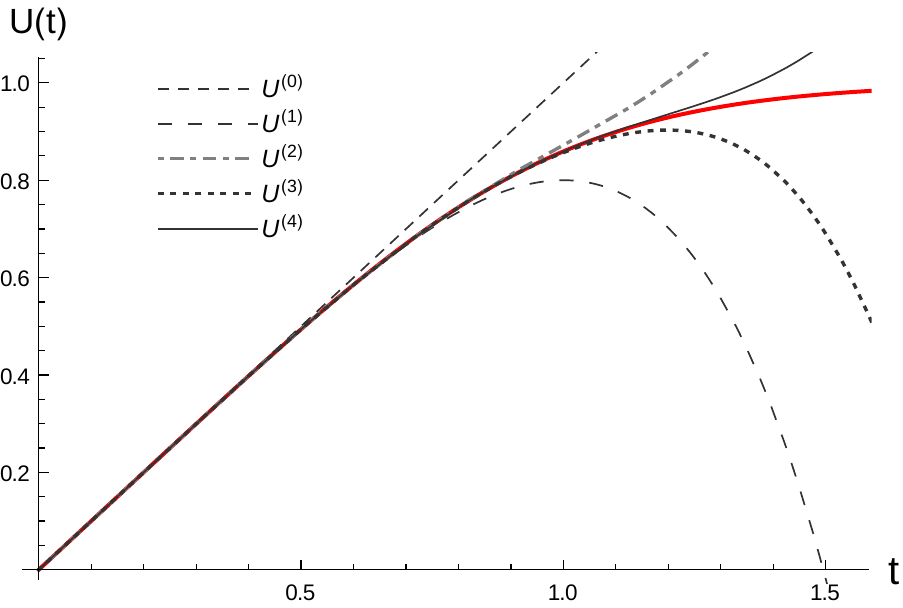}
    \caption{}
    \label{fig:F_one}
\end{subfigure}
\caption{Solution to the nonlinear FDE \eqref{eq:nonlinear_FDE} with constant source $\psi(t) = 1$ and fractional order \textbf{(a)} $\alpha = 3/4$ and \textbf{(b)} $\alpha = 1$. The numerical solution (red, full line) is compared with the approximants up to fourth order (black/gray lines).}
\label{fig:F_threequarter_one}
\end{figure}

We obtain
\begin{equation}
    \label{eq:FDE_convolution}
    U^{(0)}(t) = \int\limits_0^t G(t,t')\psi(t')\mathrm{d}t' = \frac{1}{\Gamma(\alpha)}\int\limits_0^t (t-t')^{\alpha-1}\mathrm{d}t' = \frac{t^\alpha}{\alpha\Gamma(\alpha)}
\end{equation}
The residual operator is the difference between the operators of the linear FDE \eqref{eq:linear_FDE} and the nonlinear FDE \eqref{eq:nonlinear_FDE} and is defined by the action on $U(t)$, i.e.,
\begin{equation}
    \label{eq:FDE_residual}
    \mathcal{R}_t U = \mathcal{L}_t U - \mathcal{N}_t U = -U^n
\end{equation}
Now the $p-$th order approximant to equation \eqref{eq:nonlinear_FDE} can be calculated by using the BLUES iteration sequence \eqref{eq:nth_order}
\begin{equation}
    \label{eq:FDE_iteration}
    U^{(p)}(t) = U^{(0)}(t) + \int\limits_0^t G(t,t')\mathcal{R}_{t'} U^{(p-1)}(t')\mathrm{d}t'
\end{equation}
The first-order approximant to the nonlinear problem can easily be calculated using \eqref{eq:Green_FDE}, \eqref{eq:FDE_convolution} and the iteration sequence definition \eqref{eq:FDE_iteration}, with the choice $n = 4$, 
\begin{equation}
    \label{eq:FDE_first_order}
    \begin{split}
    U^{(1)}(t) &= U^{(0)}(t) - \int\limits_0^t \frac{(t-t')^{\alpha-1}}{\Gamma(\alpha)}\frac{t'^{4\alpha}}{\alpha^4 \Gamma^4(\alpha)}\mathrm{d}t'\\
    &= \frac{1}{\alpha\Gamma(\alpha)}t^\alpha - \frac{\Gamma(1+4\alpha)}{\alpha^4 \Gamma^4(\alpha)\Gamma(1+5\alpha)}t^{5\alpha}
    \end{split}
\end{equation}
One can now iterate \eqref{eq:FDE_iteration} to generate higher-order approximants to the solution of the nonlinear FDE \eqref{eq:nonlinear_FDE}. In Fig. \ref{fig:F_quarter_half} and Fig. \ref{fig:F_threequarter_one}, the approximants for different values of $\alpha$ are compared with the numerically exact solution.

\begin{figure}[!ht]
\centering
    \includegraphics[width = 0.85\linewidth]{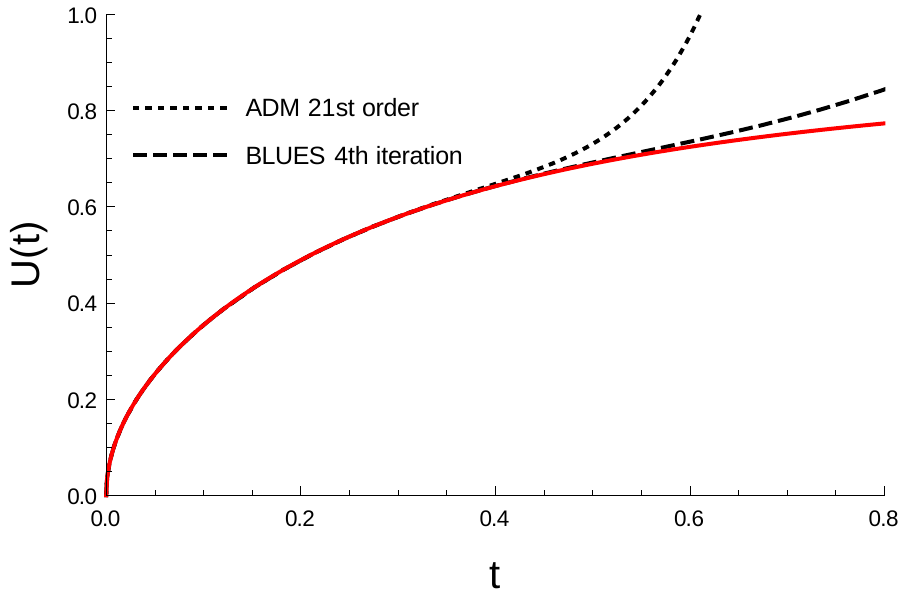}
\caption{Comparison between the 21st-order approximant of the ADM (dotted line) and the 4th iteration of the BLUES function method (dashed line) for $\alpha =1/2$ and $n=4$. The numerically exact solution (red, full line) is also shown.}
 \label{fig:F_comparison}
\end{figure}

For the choice $\alpha = 1/2$, equation \eqref{eq:nonlinear_FDE} is associated with the heat transfer equations for a semi-infinite solid \cite{keller1972} with external heating $\psi(t)$ and either linear Newton (for $n=1$) or nonlinear Stefan-Boltzmann cooling (for $n = 4$). In \cite{keller1972}, it was shown that for $n\geq3$, some of the energy entering the solid will remain, while for $n\leq2$, all energy is eventually radiated away. For the remainder of this work, we will use nonlinear Stefan-Boltzmann cooling, $n = 4$. The zeroth-, first-, and second-order approximants for $\alpha = 1/2$ and $n = 4$ are, respectively, given by
\begin{equation}
    \label{eq:FDE_12_zero_first}
    \begin{split}
        U^{(0)}(t) &= 2\left(\frac{t}{\pi}\right)^{1/2}\\
        U^{(1)}(t) &= 2\left(\frac{t}{\pi}\right)^{1/2} - \frac{256}{15}\left(\frac{t}{\pi}\right)^{5/2}\\
        U^{(2)}(t) &= 2\left(\frac{t}{\pi}\right)^{1/2} - \frac{256}{15}\left(\frac{t}{\pi}\right)^{5/2} + \frac{2097152}{4725}\left(\frac{t}{\pi}\right)^{9/2} -\frac{1073741824}{225225}\left(\frac{t}{\pi}\right)^{13/2}\\
        &+ \frac{8796093022208}{369208125}\left(\frac{t}{\pi}\right)^{17/2} - \frac{2251799813685248}{49104680625}\left(\frac{t}{\pi}\right)^{21/2}
    \end{split}
\end{equation}
The approximants obtained by the BLUES function method can be compared to those obtained with the Adomian decomposition method (ADM) by making use of the following recursion relation \cite{wazwaz2012} for the coefficients $a_n$ of the solution series of the ADM,
\begin{equation}
U_{\rm ADM}(t) = \sum_{n=0}^{\infty} a_n\, t^{\alpha n},
\end{equation}
with
\begin{equation}
    \label{eq:recursion_wazwaz}
    \begin{split}
        a_0 &= 0\\
        a_1 &= \frac{1}{\Gamma(\alpha+1)}\\
        a_{n+1} &= -\frac{\Gamma(n\alpha+1)}{\Gamma(n\alpha+\alpha+1)}A_n
    \end{split}
\end{equation}
where the $A_n$, $n\geq1$ are defined as follows
\begin{equation}
    \label{eq:wazwaz_A}
    A_n = \sum_{l=0}^n\sum_{s=0}^l\sum_{k=0}^sa_{n-l}a_{l-s}a_{s-k}a_k
\end{equation}
From equations \eqref{eq:recursion_wazwaz} and \eqref{eq:wazwaz_A} it can be deduced that for $\alpha = 1/2$ one has $a_n = 0$ for $n \neq 4k+1$, with $k=0,1,2, ...$. We will henceforth define the $p$-th order ADM approximant as the truncated series
\begin{equation}
U_{\rm ADM}^{(p)}(t) \equiv \sum_{n=0}^{p} a_n\, t^{\alpha n},
\end{equation}
A comparison is now made between BLUES and ADM. In Fig.\ref{fig:F_comparison} the 21st-order approximant for the ADM (containing five nontrivial exact terms) is shown, together with the 4th approximant for the BLUES function method (containing many more terms, but also five nontrivial exact ones -- see further) and the numerically exact solution.

The number of nonzero terms $g(i,n)$ for the $i$-th approximant generated by the BLUES function method can be calculated for a general nonlinearity exponent $n\in\mathbb{N}$, $n\geq1$, i.e.,
\begin{equation}
    \label{eq:FDE_number_of_terms}
    g(i,n) = \begin{cases}
    \frac{1}{n-1}\left(n-2+n^i\right) & n\geq 2\\
    i+1 & n =1
    \end{cases}
\end{equation}
so for $n =4$, the 4th approximant already contains  $g(4,4) = 86$ nonzero terms. Note that the number of terms in BLUES increases exponentially with the order of iteration. In comparison, the ADM generates a series with a number of terms which grows linearly with the order of the approximation. Note that the ADM generates the exact coefficients in a series expansion of the solution while the BLUES function method does not. In contrast, BLUES generates many more terms, of which only the lower-order ones are exact. The higher-order coefficients have not yet settled or converged to their exact value. We find empirically that in the BLUES function method in iteration $p$ the first $p+1$ coefficients are exact, which is a linear progression like in ADM. For example, in \eqref{eq:FDE_12_zero_first} the expressions for $U^{(0)}(t)$ and $U^{(1)}(t)$ contain the exact coefficients, while only the first three terms in $U^{(2)}(t)$ are exact. Although the higher-order terms are not exact, it appears that their presence (in BLUES) leads to a more accurate approximant than their absence (in ADM), when the comparison is made with equal numbers of exact terms. 

The BLUES function method generates, in each iteration, a (huge) number of scout terms that probe the emerging series expansion and gradually gain precision. In this respect, the BLUES function method is reminiscent in spirit of a Pad\'e approximation applied to a series expansion. The coefficients $a_n$ in the BLUES function method saturate roughly linearly with increasing order of iteration, as can be seen, e.g., when keeping track of the coefficient $a_{17}$ of the $t^{17/2}$ term. This coefficient is first generated in the second approximant $U^{(2)}$ with a provisional value of $a_{17} = 1.41659$. In the third approximant $U^{(3)}$ the value of this coefficient increases to $a_{17} = 19.8026$ and settles in the fourth approximant $U^{(4)}$, attaining its exact value $a_{17} = 30.8436$, which is the same value as is found with the ADM. Similar observations can be made for the coefficient $a_{21}$ of the $t^{21/2}$ term. This term is first generated in the second approximant with a value of $a_{21} = -0.27627$ and increases (in absolute value) to $a_{21} = -40.9762$ in the third iteration, then jumps to $a_{21} = -99.0372$ in the fourth iteration and settles at $a_{21} = -118.387$ in the fifth iteration, which is the exact result as found by the ADM.

\begin{figure}[!ht]
\centering
    \includegraphics[width = 0.85 \linewidth]{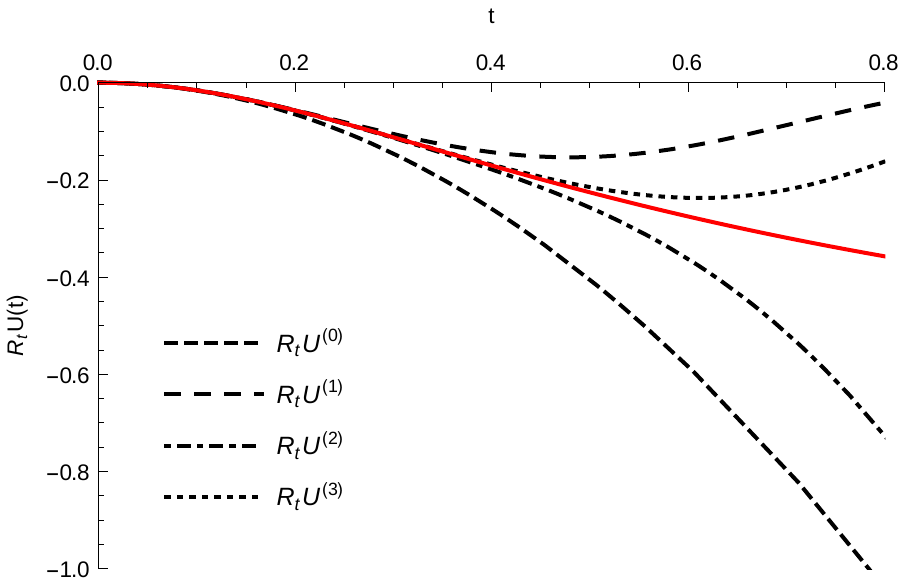}
\caption{Residual $\mathcal{R}_t U(t) = -U^4(t)$ for $\alpha = 1/2$ for different orders of iteration. The numerically exact residual $\mathcal{R}_t U_{num}$ is also shown (red, full line).}
 \label{fig:F_res_half}
\end{figure}

The higher-order residual functions are shown in Fig.\ref{fig:F_res_half} together with the residual operator \eqref{eq:FDE_residual} applied to the numerically exact solution (red, full line). Note that for this model the residuals are not localized, in contrast to all previous examples. The approximants to the solution of equation \eqref{eq:nonlinear_FDE} do not converge to the correct numerical value at $t\rightarrow\infty$ owing to the divergence of the residual in every order of iteration. While the approximants diverge for larger values of $t$, a radius of convergence can still be identified.

\section{Fisher equation}
\label{sec:fisher}
As a starting point for our final example, consider the diffusion equation which describes the propagation of a density $u(x,t)$
\begin{equation}
    \label{eq:diffusion}
    u_t - \nu u_{xx} = 0,
\end{equation}
with $\nu$ the (dimensionless) diffusion coefficient. Adopting a traveling-wave Ansatz $z = x-ct$, the diffusion equation becomes a linear ordinary DE. We add a co-moving Dirac delta source, 
\begin{equation}
  \label{eq:burgers_linear}
\mathcal{L}_zU \equiv  - U_z - k U_{zz} = \delta(z),
\end{equation}
where $k$ is a dimensionless constant. We consider the wavefront boundary conditions  $U_z(z\rightarrow - \infty) = 0$ (and $U(z\rightarrow - \infty) > 0$) and $U(z\rightarrow\infty) = 0$.
The exact solution (in every point including $z=0$) is the piecewise analytic exponential tail,
\begin{equation}
    \label{eq:burgers_BLUES}
    B(z) = \begin{cases}
    1, & z<0 \\
    \mathrm{e}^{-z/k}, & z\geq0
    \end{cases}
\end{equation}
and the wavefront velocity is $c(k) =1/k$. We now add a reaction-type nonlinearity (growth term) and a source $\psi(z)$ to obtain the forced Fisher equation \cite{fisher1937} in co-moving coordinates, i.e.,
\begin{equation}
    \label{eq:fisher}
    \nlop U = - U_z - kU_{zz} - kU(1-U) = \psi(z)
\end{equation}
with boundary conditions $U(z\rightarrow\infty) \rightarrow0$ and $U(z\rightarrow-\infty)\rightarrow1$. Equation \eqref{eq:fisher} governs the dimensionless density of some bio-chemical substance or biological population experiencing diffusion and growth. The limit at negative infinity signifies the saturation of the density at the normalized value 1. The residual operator is now acting as follows
\begin{equation}
    \label{eq:fisher_residual}
    \rop U = k U(1-U)
\end{equation}
If we now choose  $\psi(z)$ to be the exponential corner source \eqref{eq:source}, the zeroth-order approximant to the nonlinear DE \eqref{eq:fisher} is the same as was calculated for the Burgers equation in \cite{Berx}, and is repeated here in appendix \ref{app:fisher}. Higher orders can easily be calculated by iteration but will not be given here. We will, however, present the  calculated first-order approximant for $k\neq K$ in the appendix. Note that the first-order approximant approaches a constant which is not unity at negative infinity for $z$ and consequently does not obey the boundary condition. One can calculate the non-trivial constant by considering the limit of the first-order approximant at negative infinite $z$, i.e.,
\begin{equation}
    \label{eq:fisher_first_constant}
    \begin{split}
    U_c &\equiv \lim_{z\rightarrow-\infty} U^{(1)}(z) = 1 + \lim_{z\rightarrow-\infty}\int\limits_{-\infty}^\infty B(z-z') \mathcal{R}_{z'} U^{(0)}_\psi(z')\mathrm{d}z'\\
    &= 1+\int\limits_{-\infty}^\infty \mathcal{R}_{z'} U^{(0)}_\psi(z')\mathrm{d}z'\\
    &= 1+\frac{k \left(2 k^3+4 k^2 K+6 k K^2+3 K^3\right)}{4 (k+K)^2}
    \end{split}
\end{equation}

\begin{figure}[!htp]
\begin{subfigure}{\linewidth}
    \centering
    \includegraphics[width = 0.73 \linewidth]{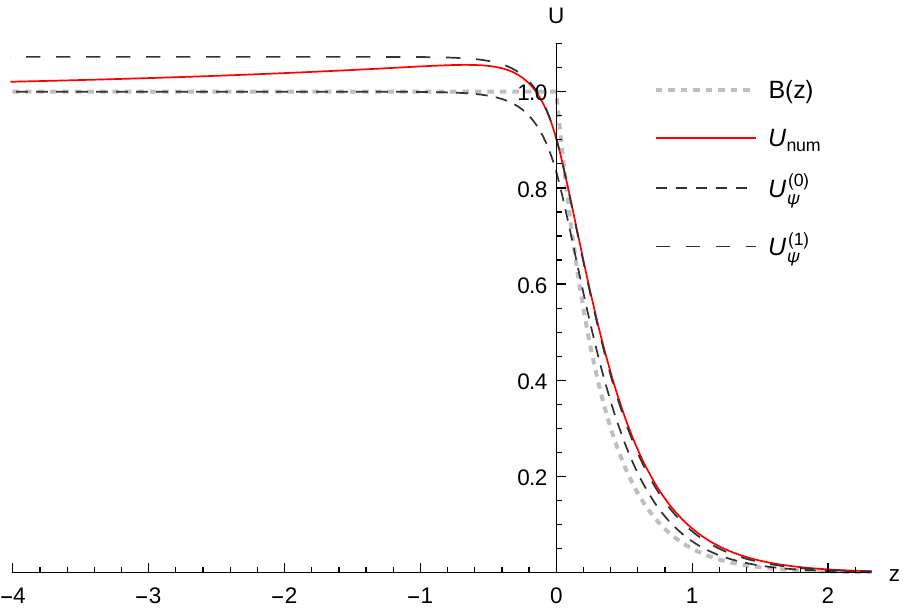}
    \caption{}
    \label{fig:Fi}
\end{subfigure}
\begin{subfigure}{\linewidth}
    \centering
    \includegraphics[width = 0.73 \linewidth]{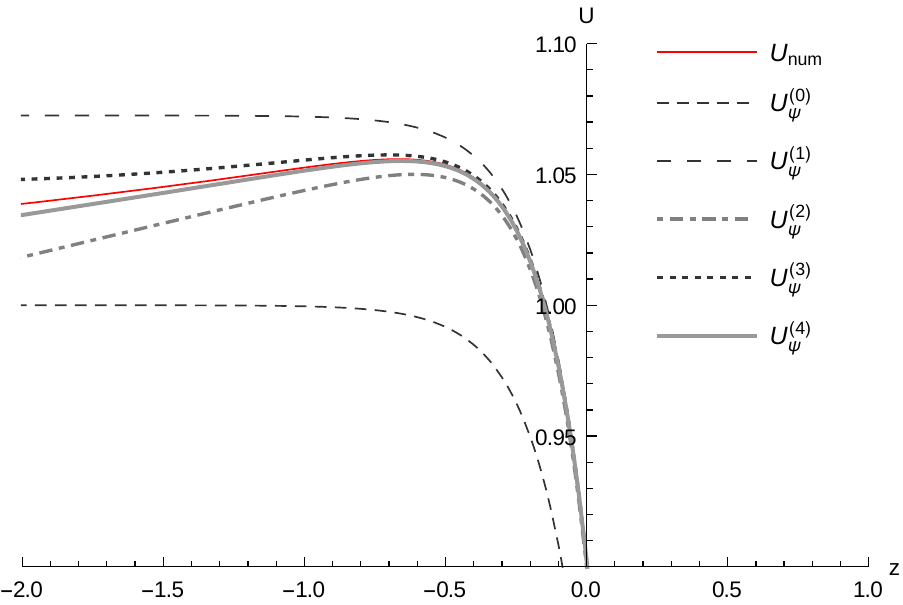}
    \caption{}
    \label{fig:Fi_zoom}
\end{subfigure}
\caption{\textbf{(a)} Traveling wavefront solution to the nonlinear Fisher DE \eqref{eq:fisher} with an exponential corner source \eqref{eq:source}. The numerical solution $U_{\rm num}$ (red, full line), the zeroth-order $U_\psi^{(0)}$ (black, dashed line) and the first-order approximant $U_\psi^{(1)}$ (black, wider spaced dashed line) are compared. The BLUES function (gray) is also shown. \textbf{(b)} A zoomed-in view around the shoulder of the wavefront. The approximants are shown up to and including 4th order. The latter (gray, full line) lies just below the numerical one (red, full line). Parameter values are $k = 1/3$ and $K/k = 1/2$.}
\end{figure}

\begin{figure}[!htp]
    \centering
    \includegraphics[width = 0.85\linewidth]{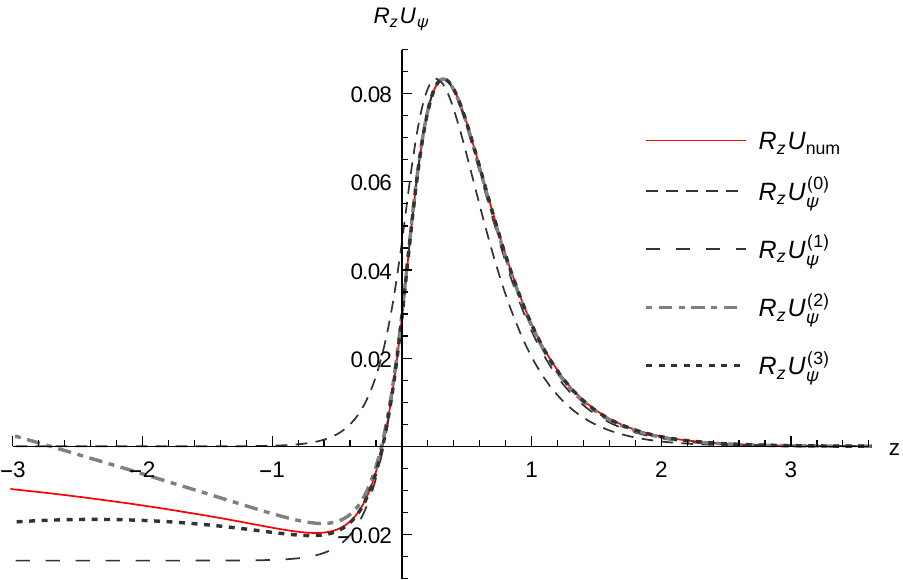}
\caption{Residual function $\rop U_\psi^{(n)}$ of the approximants of order $n = 0, 1, 2, 3$ for the nonlinear Fisher DE with exponential corner source \eqref{eq:source}. The residual operator applied to the numerical solution (red, full line) is also shown. Parameter values are $k = 1/3$ and $K/k = 1/2$.}
\label{fig:Fi_res}
\end{figure}

In Fig.\ref{fig:Fi}, the zeroth- and first-order approximants are shown together with the numerically exact solution and the BLUES function \eqref{eq:burgers_BLUES}. In Fig.\ref{fig:Fi_zoom}, a zoomed-in representation of the shoulder of the wavefront is shown. The numerical solution is compared with approximants up to fourth order. Note that while all approximants obey the boundary condition $U(z\rightarrow\infty) \rightarrow0$, only the zeroth-order approximant approaches unity for $z\rightarrow-\infty$. This is a consequence of the lack of localization of the residual for higher orders. This is shown in Fig.\ref{fig:Fi_res}. The numerical residual function $\rop U_{num}$ and the zeroth-order residual function are localized, but higher-order residual functions are not anymore. This corresponds to a divergence of the approximants of higher orders.

\begin{figure}[!ht]
    \centering
    \includegraphics[width = 0.85\linewidth]{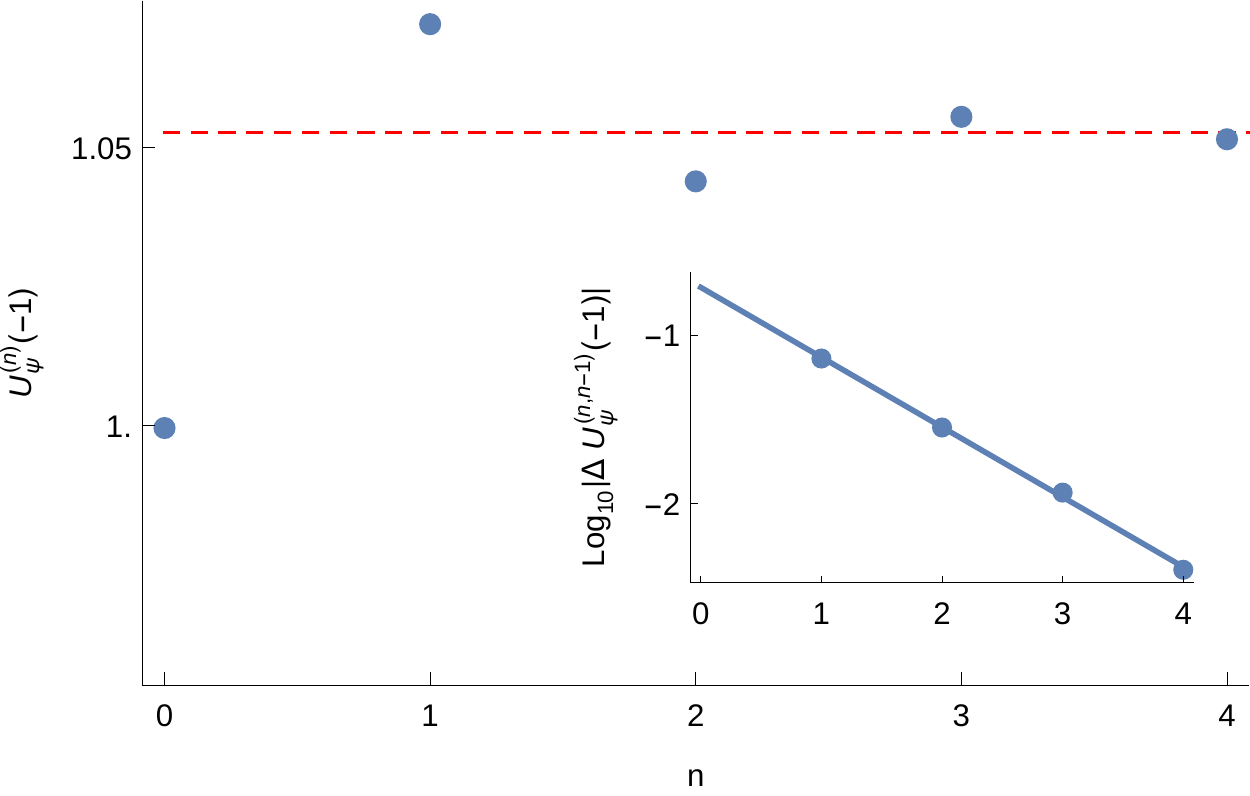}
\caption{Wavefront values $U_\psi^{(n)}(-1)$ at $z = -1$ versus order $n$ for the Fisher equation \eqref{eq:fisher}. The numerically exact value (red, dashed line) is also shown. \textbf{Inset:} A $\logten$ semi-log plot of the increments  $|\Delta U_\psi^{(n,n-1)}(-1)|$ of the approximants versus $n$, and a linear fit. These increments are defined as the difference of the $n$-th and $(n-1)$-st approximant. Parameter values are $k=1/3$ and $K/k = 1/2$.}
\label{fig:Fi_convergence}
\end{figure}

The local convergence of the approximants can nevertheless be assessed by studying the value of the approximants for a fixed value of $z$. In view of the divergence of the approximants for $z\rightarrow-\infty$, one has to be careful in choosing the value of $z$. In this case, we have opted for $z =-1$, which can be seen to lie within a reasonable region of convergence. The results are shown in Fig.\ref{fig:Fi_convergence}. Note that within the region of convergence, the approximants converge exponentially fast to the numerically exact solution. 

\section{Conclusions and outlook}
In this paper we have extended a useful and accurate approach for solving nonlinear ordinary DEs with sources, based on the BLUES function method introduced in \cite{BLUES} and first applied quantitatively in \cite{Berx}.  We have shown that the method can be applied to obtain useful approximations for the oscillating wave solution of the damped nonlinear oscillator and for traveling wavefront solutions of the Fisher equation. It can also be  extended to the arena of fractional DEs, as was shown for a heat transfer problem. In problems of wave propagation, typically a partial DE is transformed to an ordinary DE by means of a traveling wave Ansatz and the source must be assumed to be co-moving. This limitation has, however, been removed by considering a broader class of physical problems, as we have illustrated in the case of the nonlinear ordinary heat transfer DE.

In the application, in Section \ref{sec:fractional}, to the heat transfer problem we have made a detailed comparison of our iteration procedure with the Adomian decomposition method. It is found that both methods generate equal numbers of exact coefficients in a given order of the series expansion (ADM) or of the iteration (BLUES). However, the BLUES method generates an exponentially large number of approximate coefficients of higher-order terms that are not present in the ADM. The presence of these higher-order terms appears to accelerate the convergence of the approximation. 

One can think of further applications of the method.  For an overview of pertinent types of DEs for which this method could be tried, see for example \cite{Wazwaz}. We have in mind, for example, systems described by the nonlinear Schr\"odinger (or Gross-Pitaevskii) equation. This DE has been studied intensively in the past decades and is of general interest in the research on Bose-Einstein condensates, nonlinear optics and semiconductor physics \cite{Yang,Zhenya,Kippenbergeaan8083,Xuekai2017}. In the field of fluid mechanics, when studying solitary waves on shallow water, it would be interesting to see whether the BLUES function method can be used to calculate physically relevant profiles of tidal bores in the context of the recent minimal analytic model for this problem \cite{Berry2018,Berry2019}. Other opportunities arise when one considers different kinds of DEs, e.g., nonlinear partial DEs, either deterministic or stochastic \cite{KPZ,TAKEUCHI} and systems of coupled DEs \cite{Horowitz}. 

Finally, we announce that the method is also applicable to nonlinear partial differential equations (PDEs), for example in time $t$ and space $x$, which cannot be reduced to ordinary DEs. For PDEs the initial condition can be employed as the source term in the method. This is the subject of a substantial further development on which we will report in a future communication.

\appendix

\section{First-order approximant for the Fisher equation}\label{app:fisher}
The zeroth-order approximant to the solution of the nonlinear Fisher equation \eqref{eq:fisher} with exponential corner source \eqref{eq:source} is
\begin{equation}\label{eq:fisher_conv}
    U_\psi^{(0)}(z) \equiv (B\ast\psi)(z) = \frac{1}{2} \begin{cases}
        2 - \frac{K}{K+k}\mathrm{e}^{z/K}, & z<0\\ 
        \frac{K}{K-k}\mathrm{e}^{-z/K} - \frac{2k^2}{K^2 -k^2}\mathrm{e}^{-z/k}, & z\geq0,\\
    \end{cases}
\end{equation}
while for $K=k$ the convolution product results in
\begin{equation}\label{eq:fisher_conv_equal}
    U_\psi^{(0)}(z) \equiv (B\ast\psi)(z) = \begin{cases}
        1 - \frac{\mathrm{e}^{z/k}}{4}, & z<0\\ 
        \left(\frac{3}{4} + \frac{z}{2k}\right)\mathrm{e}^{-z/k}, & z\geq0.\\
    \end{cases}
\end{equation}
To calculate the first-order approximation to the solution of the Fisher equation \eqref{eq:fisher}, the residual operator \eqref{eq:fisher_residual} is applied to the zeroth-order approximant in accordance with the iteration equation \eqref{eq:nth_order}

\begin{equation}
    \label{eq:fisher_residual_on_conv}
    \rop U_\psi^{(0)}(z) =-\frac{k}{4}\begin{cases}
             \frac{K e^{z/K} \left(K e^{z/K}-2 (k+K)\right)}{(k+K)^2}, & z<0, \\
             \frac{e^{-2 z \left(\frac{1}{k}+\frac{1}{K}\right)} \left(K (k+K) e^{z/k}-2 k^2 e^{z/K}\right) \left((k+K) e^{z/k} \left(K-2 (K-k) e^{z/K}\right)-2 k^2 e^{z/K}\right)}{\left(k^2-K^2\right)^2}, & z\geq0 \\
        \end{cases}
\end{equation}
Now a convolution product of the BLUES function \eqref{eq:burgers_BLUES} with the residual \eqref{eq:fisher_residual_on_conv} is calculated. The increment between the zeroth-order and first-order approximants $\Delta U_\psi^{(1,0)} = U_\psi^{(1)} -U_\psi^{(0)}$ is presented here, which is the convolution product $\Delta U_\psi^{(1,0)}(z) = (B\ast\rop U_\psi^{(0)})(z)$
\begin{equation}
    \label{eq:fisher_first}
    \Delta U_\psi^{(1,0)}(z) =  
        \begin{cases}
         \frac{k}{8\alpha^2} \left(\frac{K^4}{2k+K}\mathrm{e}^{2z/K} -4K^3 \mathrm{e}^{z/K} + 2 (2k^3 + 4k^2 K + 6 kK^2 + 3K^3)\right), & z<0, \\
         \frac{k}{8\alpha^2\beta^2}\left(4k^5 \mathrm{e}^{-2z/k} + \frac{8k^2 K^3 (5k^2+K^2)}{\gamma}\mathrm{e}^{-z/k}-8k^2K^3\mathrm{e}^{-\alpha z/kK}\right. & \\
         \left. + 4K^3 \alpha^2 \mathrm{e}^{-z/K} + \frac{K^4 \alpha^2}{2k-K}\mathrm{e}^{-2z/K}\right), & z\geq0 ,\\
        \end{cases}
\end{equation}
where $\alpha = k + K$, $\beta = k-K$ and $\gamma = K^2 -4k^2$. The cases for which $K = k$ or $K = 2k$ must be treated separately but these (elementary) calculations will not be performed here. Note that  the first-order approximant  $U_\psi^{(1)}(z)$ approaches the constant value $U_c$ for $z\rightarrow-\infty$, which has already been presented in equation \eqref{eq:fisher_first_constant}.
\bibliography{analytic_BLUES_evolution.bib}

\begin{thebibliography}{27}%
\makeatletter
\providecommand \@ifxundefined [1]{%
 \@ifx{#1\undefined}
}%
\providecommand \@ifnum [1]{%
 \ifnum #1\expandafter \@firstoftwo
 \else \expandafter \@secondoftwo
 \fi
}%
\providecommand \@ifx [1]{%
 \ifx #1\expandafter \@firstoftwo
 \else \expandafter \@secondoftwo
 \fi
}%
\providecommand \natexlab [1]{#1}%
\providecommand \enquote  [1]{``#1''}%
\providecommand \bibnamefont  [1]{#1}%
\providecommand \bibfnamefont [1]{#1}%
\providecommand \citenamefont [1]{#1}%
\providecommand \href@noop [0]{\@secondoftwo}%
\providecommand \href [0]{\begingroup \@sanitize@url \@href}%
\providecommand \@href[1]{\@@startlink{#1}\@@href}%
\providecommand \@@href[1]{\endgroup#1\@@endlink}%
\providecommand \@sanitize@url [0]{\catcode `\\12\catcode `\$12\catcode
  `\&12\catcode `\#12\catcode `\^12\catcode `\_12\catcode `\%12\relax}%
\providecommand \@@startlink[1]{}%
\providecommand \@@endlink[0]{}%
\providecommand \url  [0]{\begingroup\@sanitize@url \@url }%
\providecommand \@url [1]{\endgroup\@href {#1}{\urlprefix }}%
\providecommand \urlprefix  [0]{URL }%
\providecommand \Eprint [0]{\href }%
\providecommand \doibase [0]{http://dx.doi.org/}%
\providecommand \selectlanguage [0]{\@gobble}%
\providecommand \bibinfo  [0]{\@secondoftwo}%
\providecommand \bibfield  [0]{\@secondoftwo}%
\providecommand \translation [1]{[#1]}%
\providecommand \BibitemOpen [0]{}%
\providecommand \bibitemStop [0]{}%
\providecommand \bibitemNoStop [0]{.\EOS\space}%
\providecommand \EOS [0]{\spacefactor3000\relax}%
\providecommand \BibitemShut  [1]{\csname bibitem#1\endcsname}%
\let\auto@bib@innerbib\@empty
\bibitem [{\citenamefont {Adomian}(1994)}]{adomian}%
  \BibitemOpen
  \bibfield  {author} {\bibinfo {author} {\bibfnamefont {G.}~\bibnamefont
  {Adomian}},\ }\href {\doibase 10.1007/978-94-015-8289-6} {\emph {\bibinfo
  {title} {Solving Frontier Problems of Physics: The Decomposition Method}}},\
  \bibinfo {edition} {1st}\ ed.,\ \bibinfo {series} {Fundamental Theories of
  Physics}, Vol.~\bibinfo {volume} {60}\ (\bibinfo  {publisher} {Springer
  Netherlands},\ \bibinfo {year} {1994})\BibitemShut {NoStop}%
\bibitem [{\citenamefont {Liao}(2012)}]{ham}%
  \BibitemOpen
  \bibfield  {author} {\bibinfo {author} {\bibfnamefont {S.}~\bibnamefont
  {Liao}},\ }\href {\doibase 10.1007/978-3-642-25132-0} {\emph {\bibinfo
  {title} {Homotopy Analysis Method in Nonlinear Differential Equations}}},\
  \bibinfo {edition} {1st}\ ed.\ (\bibinfo  {publisher} {Springer-Verlag Berlin
  Heidelberg},\ \bibinfo {year} {2012})\BibitemShut {NoStop}%
\bibitem [{\citenamefont {Karpman}\ and\ \citenamefont
  {Maslow}(1977)}]{karpman1977}%
  \BibitemOpen
  \bibfield  {author} {\bibinfo {author} {\bibfnamefont {V.~I.}\ \bibnamefont
  {Karpman}}\ and\ \bibinfo {author} {\bibfnamefont {E.~M.}\ \bibnamefont
  {Maslow}},\ }\bibfield  {title} {\enquote {\bibinfo {title} {Perturbation
  theory for solitons},}\ }\href
  {http://www.jetp.ac.ru/cgi-bin/dn/e_046_02_0281.pdf} {\bibfield  {journal}
  {\bibinfo  {journal} {J. Exp. Theor. Phys.}\ }\textbf {\bibinfo {volume}
  {73}},\ \bibinfo {pages} {537--559} (\bibinfo {year} {1977})}\BibitemShut
  {NoStop}%
\bibitem [{\citenamefont {Keener}\ and\ \citenamefont
  {McLaughlin}(1977)}]{Keener}%
  \BibitemOpen
  \bibfield  {author} {\bibinfo {author} {\bibfnamefont {J.~P.}\ \bibnamefont
  {Keener}}\ and\ \bibinfo {author} {\bibfnamefont {D.~W.}\ \bibnamefont
  {McLaughlin}},\ }\bibfield  {title} {\enquote {\bibinfo {title} {Solitons
  under perturbations},}\ }\href {\doibase 10.1103/PhysRevA.16.777} {\bibfield
  {journal} {\bibinfo  {journal} {Phys. Rev. A}\ }\textbf {\bibinfo {volume}
  {16}},\ \bibinfo {pages} {777--790} (\bibinfo {year} {1977})}\BibitemShut
  {NoStop}%
\bibitem [{\citenamefont {Maeda}\ and\ \citenamefont {Colonius}(2017)}]{MAEDA}%
  \BibitemOpen
  \bibfield  {author} {\bibinfo {author} {\bibfnamefont {K.}~\bibnamefont
  {Maeda}}\ and\ \bibinfo {author} {\bibfnamefont {T.}~\bibnamefont
  {Colonius}},\ }\bibfield  {title} {\enquote {\bibinfo {title} {A source term
  approach for generation of one-way acoustic waves in the {E}uler and
  {N}avier-{S}tokes equations},}\ }\href {\doibase
  10.1016/j.wavemoti.2017.08.004} {\bibfield  {journal} {\bibinfo  {journal}
  {Wave Motion}\ }\textbf {\bibinfo {volume} {75}},\ \bibinfo {pages} {36}
  (\bibinfo {year} {2017})}\BibitemShut {NoStop}%
\bibitem [{\citenamefont {Ermakov}\ and\ \citenamefont
  {Stepanyants}(2019)}]{Ermakov2019}%
  \BibitemOpen
  \bibfield  {author} {\bibinfo {author} {\bibfnamefont {A.}~\bibnamefont
  {Ermakov}}\ and\ \bibinfo {author} {\bibfnamefont {Y.}~\bibnamefont
  {Stepanyants}},\ }\bibfield  {title} {\enquote {\bibinfo {title} {Soliton
  interaction with external forcing within the {K}orteweg-de {V}ries
  equation},}\ }\href {\doibase 10.1063/1.5063561} {\bibfield  {journal}
  {\bibinfo  {journal} {Chaos}\ }\textbf {\bibinfo {volume} {29}},\ \bibinfo
  {pages} {013117} (\bibinfo {year} {2019})}\BibitemShut {NoStop}%
\bibitem [{\citenamefont {Berx}\ and\ \citenamefont {Indekeu}(2019)}]{Berx}%
  \BibitemOpen
  \bibfield  {author} {\bibinfo {author} {\bibfnamefont {J.}~\bibnamefont
  {Berx}}\ and\ \bibinfo {author} {\bibfnamefont {J.~O.}\ \bibnamefont
  {Indekeu}},\ }\bibfield  {title} {\enquote {\bibinfo {title} {Analytic
  iteration procedure for solitons and traveling wavefronts with sources},}\
  }\href {\doibase 10.1088/1751-8121/ab3914} {\bibfield  {journal} {\bibinfo
  {journal} {J. Phys. A-Math. Theor.}\ }\textbf {\bibinfo {volume} {52}},\
  \bibinfo {pages} {38LT01} (\bibinfo {year} {2019})}\BibitemShut {NoStop}%
\bibitem [{\citenamefont {Indekeu}\ and\ \citenamefont {Smets}(2017)}]{smets}%
  \BibitemOpen
  \bibfield  {author} {\bibinfo {author} {\bibfnamefont {J.~O.}\ \bibnamefont
  {Indekeu}}\ and\ \bibinfo {author} {\bibfnamefont {R.}~\bibnamefont
  {Smets}},\ }\bibfield  {title} {\enquote {\bibinfo {title} {Traveling
  wavefront solutions to nonlinear reaction-diffusion-convection equations},}\
  }\href {\doibase 10.1088/1751-8121/aa7a93} {\bibfield  {journal} {\bibinfo
  {journal} {J. Phys. A-Math. Theor.}\ }\textbf {\bibinfo {volume} {50}},\
  \bibinfo {pages} {315601} (\bibinfo {year} {2017})}\BibitemShut {NoStop}%
\bibitem [{\citenamefont {Indekeu}\ and\ \citenamefont
  {M\"uller-Nedebock}(2018)}]{BLUES}%
  \BibitemOpen
  \bibfield  {author} {\bibinfo {author} {\bibfnamefont {J.~O.}\ \bibnamefont
  {Indekeu}}\ and\ \bibinfo {author} {\bibfnamefont {K.~K.}\ \bibnamefont
  {M\"uller-Nedebock}},\ }\bibfield  {title} {\enquote {\bibinfo {title}
  {{BLUES} function method in computational physics},}\ }\href {\doibase
  10.1088/1751-8121/aab345} {\bibfield  {journal} {\bibinfo  {journal} {J.
  Phys. A-Math. Theor.}\ }\textbf {\bibinfo {volume} {51}},\ \bibinfo {pages}
  {165201} (\bibinfo {year} {2018})}\BibitemShut {NoStop}%
\bibitem [{\citenamefont {Ekomasov}\ \emph {et~al.}(2018)\citenamefont
  {Ekomasov}, \citenamefont {Gumerov}, \citenamefont {Kudryavtsev},
  \citenamefont {Dmitriev},\ and\ \citenamefont {Nazarov}}]{Ekomasov2018}%
  \BibitemOpen
  \bibfield  {author} {\bibinfo {author} {\bibfnamefont {E.~G.}\ \bibnamefont
  {Ekomasov}}, \bibinfo {author} {\bibfnamefont {A.~M.}\ \bibnamefont
  {Gumerov}}, \bibinfo {author} {\bibfnamefont {R.~V.}\ \bibnamefont
  {Kudryavtsev}}, \bibinfo {author} {\bibfnamefont {S.~V.}\ \bibnamefont
  {Dmitriev}}, \ and\ \bibinfo {author} {\bibfnamefont {V.~N.}\ \bibnamefont
  {Nazarov}},\ }\bibfield  {title} {\enquote {\bibinfo {title} {Multisoliton
  dynamics in the {S}ine-{G}ordon model with two point impurities},}\ }\href
  {\doibase 10.1007/s13538-018-0606-4} {\bibfield  {journal} {\bibinfo
  {journal} {Braz. J. Phys.}\ }\textbf {\bibinfo {volume} {48}},\ \bibinfo
  {pages} {576} (\bibinfo {year} {2018})}\BibitemShut {NoStop}%
\bibitem [{\citenamefont {Gul}\ \emph {et~al.}(2018)\citenamefont {Gul},
  \citenamefont {Ali},\ and\ \citenamefont {Ullah}}]{Gul}%
  \BibitemOpen
  \bibfield  {author} {\bibinfo {author} {\bibfnamefont {Z.}~\bibnamefont
  {Gul}}, \bibinfo {author} {\bibfnamefont {A.}~\bibnamefont {Ali}}, \ and\
  \bibinfo {author} {\bibfnamefont {A.}~\bibnamefont {Ullah}},\ }\bibfield
  {title} {\enquote {\bibinfo {title} {Localized modes in parametrically driven
  long {J}osephson junctions with a double-well potential},}\ }\href {\doibase
  10.1088/1751-8121/aae951} {\bibfield  {journal} {\bibinfo  {journal} {J.
  Phys. A-Math. Theor.}\ }\textbf {\bibinfo {volume} {52}},\ \bibinfo {pages}
  {015203} (\bibinfo {year} {2018})}\BibitemShut {NoStop}%
\bibitem [{\citenamefont {Starodub}\ and\ \citenamefont
  {Zolotaryuk}(2019)}]{starodub}%
  \BibitemOpen
  \bibfield  {author} {\bibinfo {author} {\bibfnamefont {I.~O.}\ \bibnamefont
  {Starodub}}\ and\ \bibinfo {author} {\bibfnamefont {Y.}~\bibnamefont
  {Zolotaryuk}},\ }\bibfield  {title} {\enquote {\bibinfo {title} {Fluxon
  interaction with the finite-size dipole impurity},}\ }\href {\doibase
  https://doi.org/10.1016/j.physleta.2019.01.051} {\bibfield  {journal}
  {\bibinfo  {journal} {Phys. Lett. A}\ }\textbf {\bibinfo {volume} {383}},\
  \bibinfo {pages} {1419 -- 1426} (\bibinfo {year} {2019})}\BibitemShut
  {NoStop}%
\bibitem [{\citenamefont {Duan}\ \emph {et~al.}(2012)\citenamefont {Duan},
  \citenamefont {Rach}, \citenamefont {Baleanu},\ and\ \citenamefont
  {Wazwaz}}]{wazwaz2012}%
  \BibitemOpen
  \bibfield  {author} {\bibinfo {author} {\bibfnamefont {J.-S.}\ \bibnamefont
  {Duan}}, \bibinfo {author} {\bibfnamefont {R.}~\bibnamefont {Rach}}, \bibinfo
  {author} {\bibfnamefont {D.}~\bibnamefont {Baleanu}}, \ and\ \bibinfo
  {author} {\bibfnamefont {A.}~\bibnamefont {Wazwaz}},\ }\bibfield  {title}
  {\enquote {\bibinfo {title} {A review of the {A}domian decomposition method
  and its applications to fractional differential equations},}\ }\href@noop {}
  {\bibfield  {journal} {\bibinfo  {journal} {Commun. Frac. Calc.}\ }\textbf
  {\bibinfo {volume} {3}},\ \bibinfo {pages} {73--99} (\bibinfo {year}
  {2012})}\BibitemShut {NoStop}%
\bibitem [{\citenamefont {Wazwaz}\ and\ \citenamefont
  {Khuri}(1996)}]{wazwaz1996}%
  \BibitemOpen
  \bibfield  {author} {\bibinfo {author} {\bibfnamefont {A.}~\bibnamefont
  {Wazwaz}}\ and\ \bibinfo {author} {\bibfnamefont {S.}~\bibnamefont {Khuri}},\
  }\bibfield  {title} {\enquote {\bibinfo {title} {A reliable technique for
  solving the weakly singular second-kind {V}olterra-type integral
  equations},}\ }\href {\doibase https://doi.org/10.1016/0096-3003(95)00279-0}
  {\bibfield  {journal} {\bibinfo  {journal} {Appl. Math. Comput.}\ }\textbf
  {\bibinfo {volume} {80}},\ \bibinfo {pages} {287 -- 299} (\bibinfo {year}
  {1996})}\BibitemShut {NoStop}%
\bibitem [{\citenamefont {Padmavally}(1958)}]{padamavally}%
  \BibitemOpen
  \bibfield  {author} {\bibinfo {author} {\bibfnamefont {K.}~\bibnamefont
  {Padmavally}},\ }\bibfield  {title} {\enquote {\bibinfo {title} {On a
  non-linear integral equation},}\ }\href
  {http://www.jstor.org/stable/24900521} {\bibfield  {journal} {\bibinfo
  {journal} {J. Math. Mech.}\ }\textbf {\bibinfo {volume} {7}},\ \bibinfo
  {pages} {533--555} (\bibinfo {year} {1958})}\BibitemShut {NoStop}%
\bibitem [{\citenamefont {Keller}\ and\ \citenamefont
  {Olmstead}(1972)}]{keller1972}%
  \BibitemOpen
  \bibfield  {author} {\bibinfo {author} {\bibfnamefont {J.}~\bibnamefont
  {Keller}}\ and\ \bibinfo {author} {\bibfnamefont {W.}~\bibnamefont
  {Olmstead}},\ }\bibfield  {title} {\enquote {\bibinfo {title} {Temperature of
  a nonlinearly radiating semi-infinite solid},}\ }\href {\doibase
  https://doi.org/10.1090/qam/403430} {\bibfield  {journal} {\bibinfo
  {journal} {Q. Appl. Math.}\ }\textbf {\bibinfo {volume} {29}},\ \bibinfo
  {pages} {559--566} (\bibinfo {year} {1972})}\BibitemShut {NoStop}%
\bibitem [{\citenamefont {Fisher}(1937)}]{fisher1937}%
  \BibitemOpen
  \bibfield  {author} {\bibinfo {author} {\bibfnamefont {R.}~\bibnamefont
  {Fisher}},\ }\bibfield  {title} {\enquote {\bibinfo {title} {The wave of
  advance of advantageous genes},}\ }\href {\doibase
  10.1111/j.1469-1809.1937.tb02153.x} {\bibfield  {journal} {\bibinfo
  {journal} {Ann. Eugen.}\ }\textbf {\bibinfo {volume} {7}},\ \bibinfo {pages}
  {355--369} (\bibinfo {year} {1937})}\BibitemShut {NoStop}%
\bibitem [{\citenamefont {Wazwaz}(2009)}]{Wazwaz}%
  \BibitemOpen
  \bibfield  {author} {\bibinfo {author} {\bibfnamefont {A.-M.}\ \bibnamefont
  {Wazwaz}},\ }\href {\doibase 10.1007/978-3-642-00251-9} {\emph {\bibinfo
  {title} {Partial Differential Equations and Solitary Waves Theory}}},\
  Nonlinear Physical Science\ (\bibinfo  {publisher} {Springer Berlin
  Heidelberg},\ \bibinfo {address} {Berlin, Heidelberg},\ \bibinfo {year}
  {2009})\BibitemShut {NoStop}%
\bibitem [{\citenamefont {Yang}\ \emph {et~al.}(2015)\citenamefont {Yang},
  \citenamefont {Yan},\ and\ \citenamefont {Mihalache}}]{Yang}%
  \BibitemOpen
  \bibfield  {author} {\bibinfo {author} {\bibfnamefont {Y.}~\bibnamefont
  {Yang}}, \bibinfo {author} {\bibfnamefont {Z.}~\bibnamefont {Yan}}, \ and\
  \bibinfo {author} {\bibfnamefont {D.}~\bibnamefont {Mihalache}},\ }\bibfield
  {title} {\enquote {\bibinfo {title} {Controlling temporal solitary waves in
  the generalized inhomogeneous coupled nonlinear {S}chr{\"o}dinger equations
  with varying source terms},}\ }\href {\doibase 10.1063/1.4921641} {\bibfield
  {journal} {\bibinfo  {journal} {J. Math. Phys.}\ }\textbf {\bibinfo {volume}
  {56}},\ \bibinfo {pages} {053508} (\bibinfo {year} {2015})}\BibitemShut
  {NoStop}%
\bibitem [{\citenamefont {Yan}\ \emph {et~al.}(2011)\citenamefont {Yan},
  \citenamefont {Zhang},\ and\ \citenamefont {Liu}}]{Zhenya}%
  \BibitemOpen
  \bibfield  {author} {\bibinfo {author} {\bibfnamefont {Z.}~\bibnamefont
  {Yan}}, \bibinfo {author} {\bibfnamefont {X.-F.}\ \bibnamefont {Zhang}}, \
  and\ \bibinfo {author} {\bibfnamefont {W.~M.}\ \bibnamefont {Liu}},\
  }\bibfield  {title} {\enquote {\bibinfo {title} {Nonautonomous matter waves
  in a waveguide},}\ }\href {\doibase 10.1103/PhysRevA.84.023627} {\bibfield
  {journal} {\bibinfo  {journal} {Phys. Rev. A}\ }\textbf {\bibinfo {volume}
  {84}},\ \bibinfo {pages} {023627} (\bibinfo {year} {2011})}\BibitemShut
  {NoStop}%
\bibitem [{\citenamefont {Kippenberg}\ \emph {et~al.}(2018)\citenamefont
  {Kippenberg}, \citenamefont {Gaeta}, \citenamefont {Lipson},\ and\
  \citenamefont {Gorodetsky}}]{Kippenbergeaan8083}%
  \BibitemOpen
  \bibfield  {author} {\bibinfo {author} {\bibfnamefont {T.~J.}\ \bibnamefont
  {Kippenberg}}, \bibinfo {author} {\bibfnamefont {A.~L.}\ \bibnamefont
  {Gaeta}}, \bibinfo {author} {\bibfnamefont {M.}~\bibnamefont {Lipson}}, \
  and\ \bibinfo {author} {\bibfnamefont {M.~L.}\ \bibnamefont {Gorodetsky}},\
  }\bibfield  {title} {\enquote {\bibinfo {title} {Dissipative {K}err solitons
  in optical microresonators},}\ }\href
  {https://science.sciencemag.org/content/361/6402/eaan8083} {\bibfield
  {journal} {\bibinfo  {journal} {Science}\ }\textbf {\bibinfo {volume} {361}}
  (\bibinfo {year} {2018})}\BibitemShut {NoStop}%
\bibitem [{\citenamefont {Ma}\ \emph {et~al.}(2017)\citenamefont {Ma},
  \citenamefont {Egorov},\ and\ \citenamefont {Schumacher}}]{Xuekai2017}%
  \BibitemOpen
  \bibfield  {author} {\bibinfo {author} {\bibfnamefont {X.}~\bibnamefont
  {Ma}}, \bibinfo {author} {\bibfnamefont {O.~A.}\ \bibnamefont {Egorov}}, \
  and\ \bibinfo {author} {\bibfnamefont {S.}~\bibnamefont {Schumacher}},\
  }\bibfield  {title} {\enquote {\bibinfo {title} {Creation and manipulation of
  stable dark solitons and vortices in microcavity polariton condensates},}\
  }\href {\doibase 10.1103/PhysRevLett.118.157401} {\bibfield  {journal}
  {\bibinfo  {journal} {Phys. Rev. Lett.}\ }\textbf {\bibinfo {volume} {118}},\
  \bibinfo {pages} {157401} (\bibinfo {year} {2017})}\BibitemShut {NoStop}%
\bibitem [{\citenamefont {Berry}(2018)}]{Berry2018}%
  \BibitemOpen
  \bibfield  {author} {\bibinfo {author} {\bibfnamefont {M.~V.}\ \bibnamefont
  {Berry}},\ }\bibfield  {title} {\enquote {\bibinfo {title} {Minimal
  analytical model for undular tidal bore profile; quantum and {H}awking effect
  analogies},}\ }\href {\doibase 10.1088/1367-2630/aac285} {\bibfield
  {journal} {\bibinfo  {journal} {New J. Phys.}\ }\textbf {\bibinfo {volume}
  {20}},\ \bibinfo {pages} {053066} (\bibinfo {year} {2018})}\BibitemShut
  {NoStop}%
\bibitem [{\citenamefont {Berry}(2019)}]{Berry2019}%
  \BibitemOpen
  \bibfield  {author} {\bibinfo {author} {\bibfnamefont {M.~V.}\ \bibnamefont
  {Berry}},\ }\bibfield  {title} {\enquote {\bibinfo {title} {Minimal model for
  tidal bore revisited},}\ }\href {\doibase 10.1088/1367-2630/ab2b19}
  {\bibfield  {journal} {\bibinfo  {journal} {New J. Phys}\ }\textbf {\bibinfo
  {volume} {21}},\ \bibinfo {pages} {073021} (\bibinfo {year}
  {2019})}\BibitemShut {NoStop}%
\bibitem [{\citenamefont {Kardar}\ \emph {et~al.}(1986)\citenamefont {Kardar},
  \citenamefont {Parisi},\ and\ \citenamefont {Zhang}}]{KPZ}%
  \BibitemOpen
  \bibfield  {author} {\bibinfo {author} {\bibfnamefont {M.}~\bibnamefont
  {Kardar}}, \bibinfo {author} {\bibfnamefont {G.}~\bibnamefont {Parisi}}, \
  and\ \bibinfo {author} {\bibfnamefont {Y.-C.}\ \bibnamefont {Zhang}},\
  }\bibfield  {title} {\enquote {\bibinfo {title} {Dynamic scaling of growing
  interfaces},}\ }\href {\doibase 10.1103/PhysRevLett.56.889} {\bibfield
  {journal} {\bibinfo  {journal} {Phys. Rev. Lett.}\ }\textbf {\bibinfo
  {volume} {56}},\ \bibinfo {pages} {889--892} (\bibinfo {year}
  {1986})}\BibitemShut {NoStop}%
\bibitem [{\citenamefont {Takeuchi}(2018)}]{TAKEUCHI}%
  \BibitemOpen
  \bibfield  {author} {\bibinfo {author} {\bibfnamefont {K.~A.}\ \bibnamefont
  {Takeuchi}},\ }\bibfield  {title} {\enquote {\bibinfo {title} {An appetizer
  to modern developments on the {K}ardar-{P}arisi-{Z}hang universality
  class},}\ }\href {\doibase 10.1016/j.physa.2018.03.009} {\bibfield  {journal}
  {\bibinfo  {journal} {Physica A}\ }\textbf {\bibinfo {volume} {504}},\
  \bibinfo {pages} {77} (\bibinfo {year} {2018})}\BibitemShut {NoStop}%
\bibitem [{\citenamefont {Horowitz}\ and\ \citenamefont
  {Kardar}(2019)}]{Horowitz}%
  \BibitemOpen
  \bibfield  {author} {\bibinfo {author} {\bibfnamefont {J.~M.}\ \bibnamefont
  {Horowitz}}\ and\ \bibinfo {author} {\bibfnamefont {M.}~\bibnamefont
  {Kardar}},\ }\bibfield  {title} {\enquote {\bibinfo {title} {Bacterial range
  expansions on a growing front: Roughness, fixation, and directed
  percolation},}\ }\href {\doibase 10.1103/PhysRevE.99.042134} {\bibfield
  {journal} {\bibinfo  {journal} {Phys. Rev. E}\ }\textbf {\bibinfo {volume}
  {99}},\ \bibinfo {pages} {042134} (\bibinfo {year} {2019})}\BibitemShut
  {NoStop}%
\end{thebibliography}%
\end{document}